\documentclass[final,5p,times,twocolumn]{elsarticle}

\usepackage{amsmath,amssymb}
\usepackage{braket}
\usepackage[hidelinks]{hyperref}
\biboptions{sort&compress}

\newcommand{\dint}{{\rm d}}
\newcommand{\xpom}{x_{I\!P}}

\journal{Physics Letters B}

\begin{document}

\begin{frontmatter}

\title{Measuring Dual-Superconductor Length Scales in the QCD Vacuum at HERA}

\author[iitd]{Tobias Toll\corref{cor1}}
\ead{tobiastoll@iitd.ac.in}
\author[iitd]{Nahid Vasim}
\cortext[cor1]{Corresponding author}

\affiliation[iitd]{organization={Department of Physics, Indian Institute of Technology Delhi},
            addressline={Hauz Khas},
            city={New Delhi},
            postcode={110016},
            country={India}}

\begin{abstract}
In the dual-superconductor picture, the QCD vacuum behaves like a superconductor for chromoelectric fields, with characteristic penetration and coherence lengths. These scales have so far been studied primarily through lattice calculations of static flux tubes. We show that they are also imprinted on exclusive diffraction, where the transverse profile of a gluonic hotspot reflects the response of the vacuum to a localised colour source. Using the Clem vortex profile, we fit all 104 H1 and ZEUS measurements of exclusive $J/\psi$ photoproduction and find $\chi^2/{\rm ndf}=0.72$, the penetration depth $\lambda=0.19(2)~{\rm fm}$, the hotspot core size $\xi_v=0.087(14)~{\rm fm}$, and the Ginzburg--Landau parameter $\kappa=2.6(8)$, placing the effective vacuum description in the type-II regime. The corresponding vector mass $1/\lambda=1.03(11)~{\rm GeV}$ agrees with the lattice gluelump mass, and the extracted parameters reproduce the QCD string tension.
\end{abstract}


\end{frontmatter}
\section{Introduction}
\label{sec:intro}
Due to the gluon's self-interaction, the vacuum of quantum chromodynamics (QCD) contains condensates. One candidate, proposed half a century ago by 't Hooft and Mandelstam~\cite{tHooft:1981bkw,Mandelstam:1974pi}, is a dual superconductor which is similar to an ordinary superconductor with electricity and magnetism interchanged, a condensate where chromomagnetic monopoles play the role of Cooper pairs. 

A superconductor is a condensate of electron Cooper pairs that expels an external magnetic field. If the field is strong enough, a type-II superconductor admits magnetic flux through thin flux tubes that repel one another, while a type-I superconductor instead forms non-superconducting domains. At the centre of a type-II flux tube, the condensate is suppressed to zero. Two lengths characterise the condensate and fix the cross-sectional shape of the flux tubes: the coherence length, over which the condensate recovers at the centre, which is the size of the tube's core, and the penetration depth, over which the field is screened. The profile can be obtained by minimising the Ginzburg–Landau (GL) free energy of the condensate's response to the magnetic flux. While there is no exact analytic solution to the GL equations, Clem suggested a very good approximate flux-tube profile \cite{Clem:1975ohd} in 1975.

In QCD, quarks and gluons act as sources of the chromoelectric field, and the condensate as a dual superconductor expels and squeezes the field into flux tubes. This is a mechanism for confinement: the chromoelectric field binding two colour charges is confined into a tube with finite transverse size, whose energy grows linearly with their separation, until it becomes favourable to create a new quark–antiquark pair, producing colour-neutral hadrons. An isolated quark or gluon is therefore never observed. In this picture, the QCD vacuum is then characterised by its own coherence length $\xi$ and penetration depth $\lambda$, whose ratio, the GL parameter $\kappa=\lambda/\xi$, separates type-I from type-II behaviour at $\kappa=1/\sqrt2$ and determines whether flux tubes are stable and repel.

These parameters have been studied extensively on the lattice, through field strength correlators~\cite{DiGiacomo:1992src,DElia:1997ne} and through the transverse profiles of static flux tubes~\cite{Cea:2012qw,Cea:2014uja,Shibata:2012ae,Shibata:2019nit}, including their entanglement entropy in 2+1 dimensions~\cite{Amorosso:2024jhep,Amorosso:2026mdo}. The associated vacuum excitations have been computed as gluelump~\cite{Bali:2003jq, Berwein:2015vca} and glueball~\cite{Athenodorou:2020ani,Morningstar:1999rf} masses. They have not, to our knowledge, been extracted from scattering data\footnote{This Letter supersedes our earlier preprint~\cite{Toll:2026hmx}.}.

We argue here that they can be. At the HERA accelerator electrons and protons were colliding at centre-of-mass energies $\sqrt{s}=318~{\rm GeV}$. A high-energy collision between an electron $e$ and a proton $p$ is mediated by a virtual photon. In a subset of collisions, so-called diffractive events, no quantum numbers, in particular no colour charge, are exchanged between the virtual photon and the proton, which either stays intact or dissociates, leaving a large rapidity gap between the proton remnant and the rest of the final state. At high energy we may view the colourless exchange as being mediated by two gluons with opposite colour charges. These two gluons together carry a momentum fraction $\xpom$ of the proton's momentum. In so-called exclusive diffraction there is only one extra particle in the final state. The interaction is then $ep\rightarrow e'p'V$, or $ep\rightarrow e'YV$ when the proton dissociates into a system $Y$, where $V$ denotes a vector meson such as $\rho$, $\phi$, or $J/\psi$. Exclusive diffractive scattering measures the transverse spatial structure of the target, since the transverse momentum difference between the outgoing and incoming proton $\vec\Delta$ is Fourier conjugate to the impact parameter $\vec b$. At small $\xpom$ the proton's gluon content is concentrated into hotspots~\cite{Mantysaari:2016ykx,Mantysaari:2016jaz,Cepila:2016uku}, each of which is the gluonic cloud surrounding a localised colour source. 

In the dual-superconductor picture, a colour source triggers a response from the QCD condensate. The probe crosses the proton at fixed light-front time and thus sees a frozen snapshot of its colour field. Since the proton is a stationary state, each snapshot already contains the equilibrated response of the vacuum to the colour sources present, on the scales $\xi_v$ and $\lambda$. This motivates using the Clem vortex profile~\cite{Clem:1975ohd} as an effective parametrisation of the hotspot geometry. Fitting the profile to data from the HERA experiments H1 and ZEUS determines the core parameter $\xi_v$ and penetration depth $\lambda$. From these we can then determine the GL parameter $\kappa$ and the GL coherence length $\xi$ from the fitted shape.

Previous hotspot analyses assumed Gaussian profiles. These have two known shortcomings: they fail to describe the incoherent spectrum above $|t|\simeq2~{\rm GeV}^2$ on their own, and they fail to describe the coherent and incoherent $t$-spectra simultaneously. Various remedies have been proposed, including a Bose--Einstein-like correction~\cite{Kumar:2021zbn}, an explicit hotspot evolution~\cite{Kumar:2024kns}, and pointlike fluctuating sources within Gaussian hotspots \cite{Demirci:2022wuy,Le:2025bvq} later evolved with JIMWLK~\cite{Mantysaari:2025ltq}. We show that the Clem profile, with no substructure and no evolution, describes all available HERA $J/\psi$ photoproduction data across three decades in $|t|$.

The analytic derivation is given in~\ref{app:moments}, and the systematic studies on which our conclusions rest, including fits with alternative hotspot profiles, in~\ref{app:systematics}.

\section{Framework}
\label{sec:framework}
We work in the dipole picture~\cite{Nikolaev:1990ja,Mueller:1993rr} in which the virtual photon $\gamma^*$ fluctuates into a quark-antiquark ($q\bar q$) colour dipole which scatters off the proton before forming the vector meson. There are two classes of diffractive events: \emph{incoherent} events where the proton is excited in the interaction and subsequently de-excites by breaking up, and \emph{coherent} events where it remains intact. In the Good--Walker picture~\cite{Good:1960ba}, the coherent cross section is given by the squared first moment of the amplitude and the incoherent cross section by its variance:
\begin{eqnarray}
\frac{\dint\sigma_{\rm coh.}}{\dint t}=\frac{\big|\braket{\mathcal{A}}\big|^2}{16\pi},~~
\frac{\dint\sigma_{\rm inc.}}{\dint t}=\frac{\braket{|\mathcal{A}|^2}-\big|\braket{\mathcal{A}}\big|^2}{16\pi}
\end{eqnarray}
We use the IPnonsat dipole amplitude \cite{Kowalski:2003hm, Kowalski:2006hc} which describes the interaction between the proton and the dipole as a single two-gluon exchange. At HERA kinematics this model has been shown to describe both inclusive and exclusive data well~\cite{Mantysaari:2018nng,Lappi:2010dd,Sambasivam:2019gdd,Kumar:2024kns}. The amplitude is:
\begin{eqnarray}
    \mathcal{A}^{\gamma^*p}(\xpom, \Delta)&=&\int_0^\infty\dint r\int_0^1\dint zD_{rz}^{\xpom}
    \int\dint^2\vec b e^{-i\vec\Delta\vec b}T_p(\vec b) \nonumber
\end{eqnarray}
with
\begin{eqnarray}
    D_{rz}(\xpom, \Delta)&=&(\Psi^*\Psi_V)(r, z)\frac{r^3\pi^2}
    {4 N_C}\nonumber\\
    &\times&\alpha_S(\mu^2)\xpom g(\xpom,\mu^2)J_0\left(\left(\frac12\!-\!z\right) r\Delta\right)\nonumber
\end{eqnarray}
where $r$ is the dipole size, $z$ the photon's momentum fraction taken by the quark, and $\vec b$ the impact parameter, whose conjugate $|\Delta|=\sqrt{-t}$. We use the Boosted Gaussian wave-overlap $(\Psi^*\Psi_V)$  between the photon and the vector meson \cite{Kowalski:2006hc}, and $\xpom g$ is the DGLAP evolved gluon density. There are so-called real part and skewedness corrections to the amplitude which are included as in~\cite{Kowalski:2006hc,Shuvaev:1999ce}. The first and second moments of the amplitude are directly proportional to the first and second moments of the Fourier transform of the thickness function $\tilde T_p$, which describes the transverse geometry of the proton's gluon cloud. Because the $\vec b$-dependence factorises, both moments can be obtained analytically (\ref{app:moments}).

\begin{figure}
    \centering
    \includegraphics[width=1\linewidth]{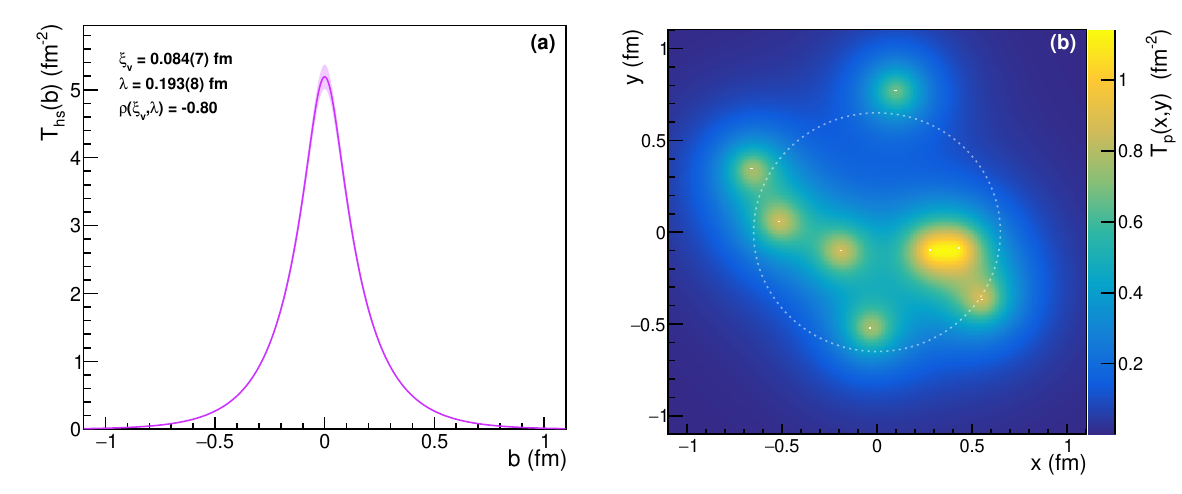}
    \caption{(a) The resulting hotspot thickness profile. (b) The proton's thickness profile at $N_{hs}=8$ and $\xpom=x_0$ using the parameter values from Table~\ref{tab:results} for one hotspot configuration. The white circle shows the proton's gluon radius $b_{\rm rms}=0.65~{\rm fm}$.}
    \label{fig:profiles}
\end{figure}

We treat the proton thickness function as a sum of $N_{hs}$ identical hotspots,
$T_p(\vec b)=1/N_{hs}\sum_i T_{hs}(\vec b-\vec b_i)$ whose centres are distributed as a Gaussian.
We can write the average and the variance of the proton's form factor as:
\begin{eqnarray}
\braket{\tilde T_p}&=&\hat T_{hs}F(t),\\
{\rm Var}_{\rm geom.}&=&\frac{\hat T_{hs}^2}{N_{hs}}\big[1+(N_{hs}-1)G(t)\big]
-\hat T_{hs}^2F^2(t)
\label{eq:var}
\end{eqnarray}
where $\hat T_{hs}(t)$ is the hotspot form factor,
$G(t)=\phi^2(t)$ the hotspot pair correlator, and
\begin{eqnarray}
F(t)=\phi\Big(\tfrac{N_{hs}-1}{N_{hs}}\Delta\Big)
\Big[\phi\Big(\tfrac{\Delta}{N_{hs}}\Big)\Big]^{N_{hs}-1}
=G(t)^{\frac{N_{hs}-1}{2N_{hs}}}
\label{eq:F}
\end{eqnarray}
is the characteristic function after shifting each configuration so that its
centre of mass (CM) lies at the origin. $\phi(t)$ is the characteristic function of the centre distribution before the CM shift. The centre-of-mass shift is required, since without it the variance is non-zero for $N_{hs}=1$. For a Gaussian hotspot centre distribution $G(t)=\exp(-B_c|t|)$, where the width $B_c$ is a free parameter. For $|t|\gg 1/B_c$, $G, F\rightarrow0$ and the geometric variance in Eq.~\eqref{eq:var} depends directly on the single-hotspot form factor.

Following~\cite{Mantysaari:2016jaz,McLerran:2015qxa} each hotspot has a log-normal saturation-scale fluctuation $\zeta_i$, normalised to $\braket{\zeta_i}=1$ so that the coherent cross section is unchanged, giving
\begin{eqnarray}
{\rm Var}_{\rm total}={\rm Var}_{\rm geom.}
+\big(e^{4\lambda_g^2\sigma_S^2}-1\big)\frac{\hat T_{hs}^2}{N_{hs}}
\label{eq:vartot}
\end{eqnarray}
with $\lambda_g=\dint\log\mathcal{A}/\dint\log(1/\xpom)$.

The analytic moments were checked against explicit Monte Carlo generation of hotspot configurations, for both moments and for all hotspot profiles used in this paper, and agree within statistical uncertainties.
\section{The Clem profile}
\label{sec:clem}
Clem's variational solution of the GL equations~\cite{Clem:1975ohd} gives a cross-sectional profile of the vortex field which we take as the hotspot thickness profile:
\begin{eqnarray}
T_{hs}(\vec b)&\propto& K_0\Bigg(\frac{\sqrt{b^2+\xi_v^2}}{\lambda}\Bigg)\nonumber\\
\hat T_{hs}(t)&=&\frac{1}{\lambda Q}\frac{K_1(\xi_v Q)}{K_1(\xi_v/\lambda)}
~~Q=\sqrt{1/\lambda^2+|t|}
\label{eq:Clem}
\end{eqnarray}
The profile has a core of size $\xi_v$ and an exponential halo of range $\lambda$. Here $\xi_v$ is the variational core parameter of Clem's ansatz which is distinct from the GL coherence length $\xi$. The GL parameter can be directly calculated from the fitted parameters~\cite{Clem:1975ohd},
\begin{eqnarray}
\kappa=\frac{\sqrt2}{\alpha}\sqrt{1-\Bigg(\frac{K_0(\alpha)}{K_1(\alpha)}\Bigg)^2},
~~\alpha=\frac{\xi_v}{\lambda}
\label{eq:GL}
\end{eqnarray}
after which $\xi=\lambda/\kappa$. Following~\cite{Nishino:2019bzb} the penetration depth and the GL coherence length determine the corresponding vector and scalar masses through $\lambda=1/m_V$ and $\xi=\sqrt2/m_S$.

The HERA data cover a wide range of photon--proton invariant mass $W$. We let the proton's width evolve as $B_c=B_{c0}+2\alpha'\log(x_0/\xpom)$ with $x_0=\xpom(W=90~{\rm GeV},t=0)=0.0012$, so that from Eq.~\eqref{eq:F} the observable shrinkage is $\alpha'_{\rm eff.}=\alpha'(N_{hs}-1)/N_{hs}$.

\begin{figure}
\centering
\includegraphics[width=1\linewidth]{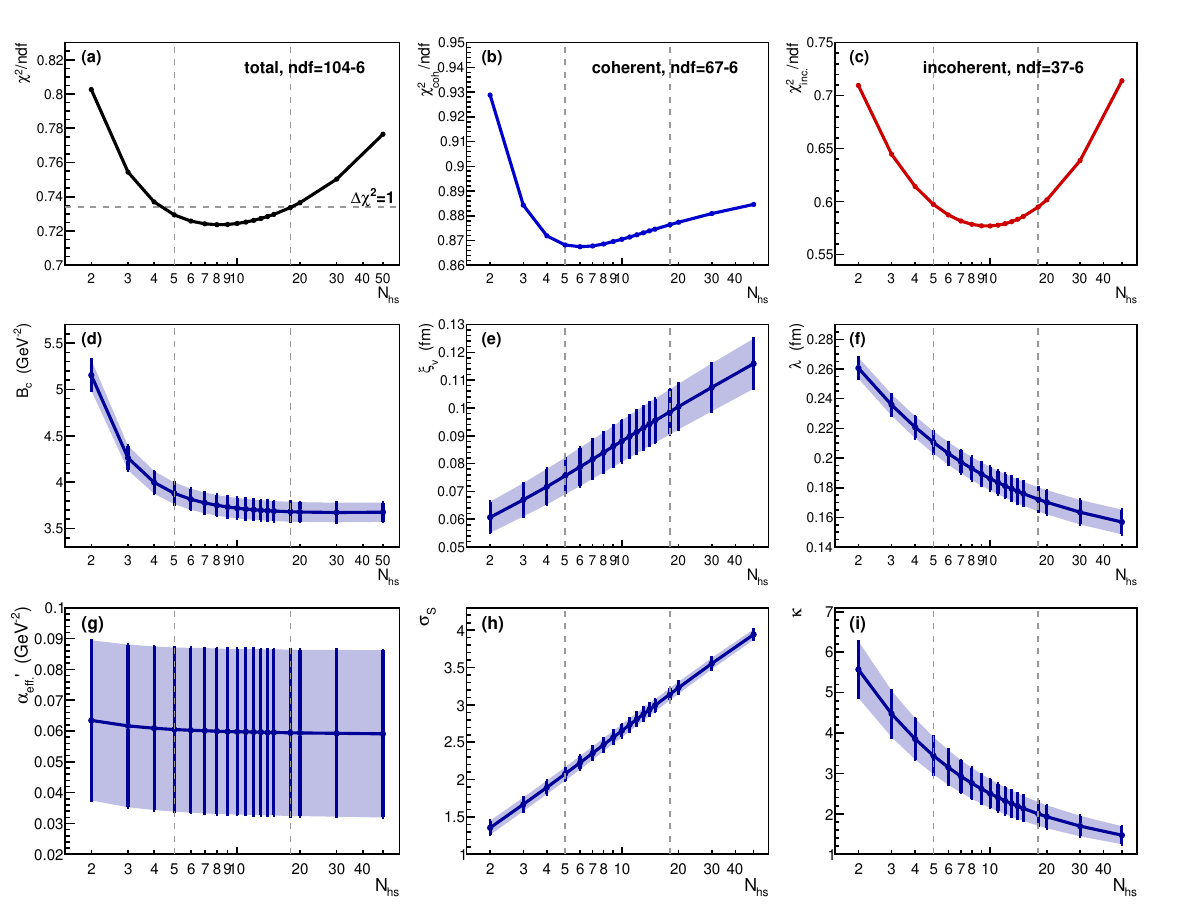}
\caption{Fit results for the Clem profile as a function of $N_{hs}$. (a) Total $\chi^2/{\rm ndf.}$, (b) the coherent $\chi^2_{\rm coh.}/{\rm ndf}$ component of the total, (c) the incoherent $\chi^2_{\rm inc.}/{\rm ndf}$ component of the total, (d) the hotspot centre distribution width $B_c$, (e) the core size parameter $\xi_v$, (f) the penetration depth $\lambda$, (g) $\alpha'_{\rm eff.}$, (h) the size of the saturation scale fluctuations $\sigma_S$, and (i) the resulting GL parameter $\kappa$.}
\label{fig:clem}
\end{figure}

\section{Comparison to HERA data}
\label{sec:data}
We fit to exclusive $J/\psi$ photoproduction, in which the virtual photon is quasi-real, from H1~\cite{Aktas:2005xu,Alexa:2013xxa,Aktas:2003zi} and ZEUS~\cite{Chekanov:2002rm,Chekanov:2009ab} with 104 points in total, 67 coherent and 37 incoherent, covering $40\leq W\leq 251~{\rm GeV}$ and $0<|t|\leq 30~{\rm GeV}^2$. This gives a good lever arm in $W$ for $\alpha'$ and in $t$ for the hotspot shape parameters. Six parameters are fitted: $B_{c0}$, $\xi_v$, $\lambda$, $\sigma_S$, $\alpha'$, and an overall normalisation. 
We use the MZ model parameters \cite{Mantysaari:2018nng}. (We note that the $J/\psi$ data in this paper have a slight preference for the MZ parametrisation over the STU one \cite{Sambasivam:2019gdd} with $\Delta\chi^2\simeq 2.7$, while $\lambda$, $\xi_v$ and $\kappa$ agree to within 1\%. The difference is that the STU dipole includes a Gaussian suppression of large dipoles.)
The normalisation comes out $\sim1.11$ and absorbs well-known sensitivities of the dipole normalisation to the charm mass and the vector-meson wavefunction, as well as normalisation uncertainties in the data sets. Being an overall factor it does not affect any extracted shape parameter. Uncertainties are obtained by varying parameters such that $\Delta\chi^2=1$. We show the resulting hotspot and proton profiles in Fig.~\ref{fig:profiles}.

Figure~\ref{fig:clem} shows the fit as a function of the number of hotspots. All the data are well described for every $N_{hs}$. The $\chi^2$ has a shallow minimum over $6\leq N_{hs}\leq 12$, with $\chi^2/{\rm ndf}=70.9/98=0.72$ at $N_{hs}=8$, and the $1\sigma$ range spans $5\leq N_{hs}\leq 18$. The coherent data prefer $6$ hotspots and the incoherent data $10$. The low $\chi^2/{\rm ndf}<1$ is consistent with other analyses, e.g. in \cite{Le:2025bvq} the authors report $\chi^2/{\rm ndf}=0.45$ with 6 parameters against a subset of the same data with $|t|<8~{\rm GeV}^2$, and could be the result of correlated systematics added in quadrature. That the HERA data do not constrain the number of hotspots well is also consistent with what was found in \cite{Le:2025bvq}.

\begin{table}
\centering
\begin{tabular}{lc}
\hline
$B_{c0}$ (GeV$^{-2}$)      & 3.75(11)  \\
$\xi_v$ (fm)               & 0.084(7)  \\
$\lambda$ (fm)             & 0.193(8)  \\
$\sigma_S$                 & 2.47(9) \\
$\alpha'_{\rm eff.}$ (GeV$^{-2}$)     & 0.060(27) \\
\hline
$\chi^2/{\rm ndf}$                       & 70.9/98=0.72  \\
$\chi^2_{\rm coh.}$ & 53.0 \\
$\chi^2_{\rm inc.}$ & 17.9  \\
\hline
\end{tabular}

\vspace{8pt}
\begin{tabular}{lrrrrr}
\hline
 & $B_{c0}$ & $\xi_v$ & $\lambda$ &
  $\sigma_{S}$ & $\alpha'$ \\
\hline
$B_{c0}$
    & 1.000 &       &        &        &       \\
$\xi_v$
    & 0.461 & 1.000 &        &        &        \\
$\lambda$
    &-0.693 &-0.799 & 1.000  &        &       \\
$\sigma_{S}$
    & -0.728 & -0.250 & 0.701 & 1.000  &       \\
$\alpha'$
    &-0.050 & 0.001 & 0.008  & 0.039  & 1.000 \\
\hline
\end{tabular}
\caption{Fit results for the Clem profile at $N_{hs}=8$ and the correlation matrix of the parameters. There are 67 coherent and 37 incoherent diffraction data points.}
\label{tab:results}
\end{table}

\begin{figure}
\centering
\includegraphics[width=1.\linewidth]{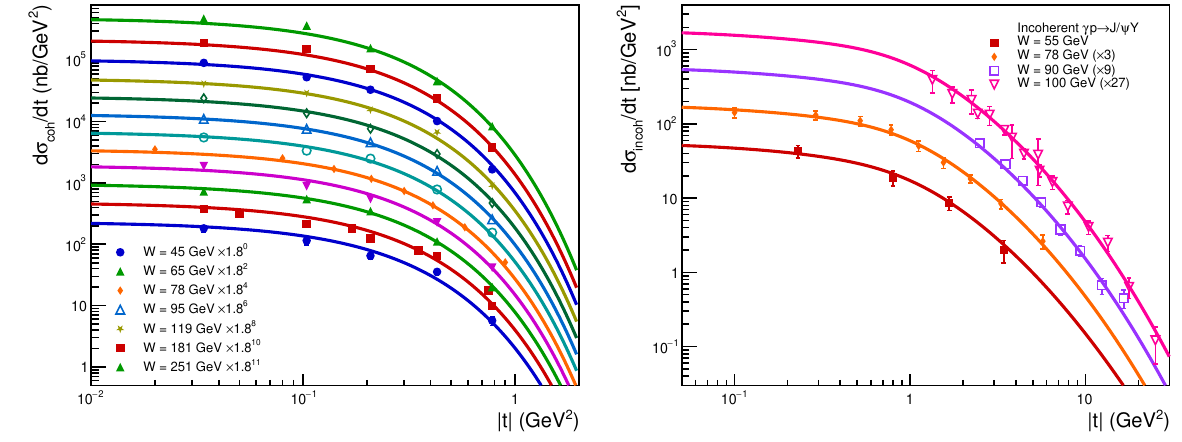}
\caption{Comparison between the fit of Table~\ref{tab:results} and coherent (left) and
incoherent (right) measurements from H1~\cite{Aktas:2005xu,Alexa:2013xxa,Aktas:2003zi} and
ZEUS~\cite{Chekanov:2002rm,Chekanov:2009ab}.}
\label{fig:data}
\end{figure}

Table~\ref{tab:results} gives the fitted parameters and their correlations, and Fig.~\ref{fig:data} the comparison to data, both at $N_{hs}=8$. The description is good across the entire $t$- and $W$-spectra.

\section{Vacuum parameters}
\label{sec:vacuum}
Across the best-fit range in $5\le N_{hs}\le 18$ we obtain:
\begin{eqnarray}
\lambda &=& 0.191(8)_{\rm fit}(19)_{N_{hs}}~{\rm fm}\nonumber\\
\xi_v &=& 0.087(7)_{\rm fit}(12)_{N_{hs}}~{\rm fm}\nonumber\\
\kappa &=& 2.6(4)_{\rm fit}(7)_{N_{hs}}
\end{eqnarray}
where the first uncertainty is from the fit and the second from the variation over $N_{hs}$.

Since $\kappa>1/\sqrt2$ by almost a factor of four, this corresponds to the type-II superconductor regime, where flux tubes are stable and repel one another. The corresponding coherence length is $\xi=\lambda/\kappa=0.073(15)~{\rm fm}$.
Lattice fits of the Clem profile to static SU(3) flux tubes by Cea et al.\ find a comparable penetration depth, $\lambda=0.175(6)$~fm (using $\sqrt{\sigma}=0.420~{\rm GeV}$, where $\sigma$ is the flux tube string tension), but a much larger core, $\xi_v\approx0.6$~fm (inferred from their $\kappa$ and $\lambda$ via Eq.~\eqref{eq:GL}), and $\kappa\approx0.2$, corresponding to a type-I vacuum~\cite{Cea:2012qw, Cea:2014uja}. Shibata et al.\ find $\kappa\approx 0.45\dots0.48$~\cite{Shibata:2012ae} which is also type-I. A static tube, however, is broadened by transverse string vibrations that grow logarithmically with the source separation~\cite{Luscher:1980iy,Cardoso:2013lla}, which a Clem fit partly absorbs into $\xi_v$. Cardoso et al.\ show that in 3+1-dimensional SU(3) the squared width grows by $0.076(9)~{\rm fm}^2$ per unit of $\ln R$ for $R=0.4\dots1.4$~fm, while the penetration depth stays at $\lambda\approx0.22\dots0.24$~fm~\cite{Cardoso:2013lla}.

The two lengths translate into vacuum excitation masses. The penetration depth gives
\begin{eqnarray}
m_V=1/\lambda=1.03(11)~{\rm GeV}
\end{eqnarray}
to be compared with the mass of the lightest ($1^{+-}$) gluelump, extracted from quenched lattice data. Since the static adjoint source results in a linearly divergent self-energy, this mass is defined only after the associated renormalon is removed, and its value depends on the scheme: $0.87(15)$~GeV in the renormalon-subtracted scheme at $\nu_f=1$~GeV~\cite{Bali:2003jq,Berwein:2015vca}, and
$2.44(7)/r_0\simeq1.0$~GeV in the scale-independent principal-value scheme~\cite{Ayala:2025trr}.

A further consistency check comes from the GL line tension. For a Clem flux tube, the energy per unit length is $\sigma=C\alpha_SL(\alpha)/\lambda^2$, where $C$ is a colour factor and $L$ the dimensionless line energy, obtained by substituting Clem's variational condition, Eq.~(16) of Ref.~\cite{Clem:1975ohd}, into his line energy, Eq.~(15) (\ref{app:string}). In the best fit range we find $L(\alpha)=1.79(22)$. The colour factor $C=|\vec\omega_k|^2=(N_c-1)/2N_c=1/3$ follows from the normalisation of the dual Ginzburg--Landau theory in Ref.~\cite{Koma:2002cv}. Athenodorou and Teper find $r_0\sqrt{\sigma_{0}}=1.160(6)$~\cite{Athenodorou:2020ani}, and Sommer gives $r_0=0.472(5)~{\rm fm}$~\cite{Sommer:2014mea} resulting in $\sqrt{\sigma_{0}}=0.485(6)~{\rm GeV}$. In order to match this we need the coupling to be $\alpha_S=0.37(4)$. 
The leading order running coupling in the $\overline{\rm MS}$ scheme that we use in our dipole model gives $\alpha_S(1/\lambda^2)=0.36(2)$ and $\alpha_S(1/\xi_v^2)=0.26(1)$, where the uncertainties reflect only those of $\lambda$ and $\xi_v$. The required value thus matches the coupling at the penetration depth. With $r_0=0.5$~fm it becomes $\alpha_S=0.33(3)$, and with $\sqrt{\sigma_0}=0.44~{\rm GeV}$ it becomes $\alpha_S=0.31(3)$, between the two scales.

The core size gives a mass $m_{\rm core}=1/\xi_v=2.3(4)~{\rm GeV}$, and the GL coherence length gives $m_S=\sqrt2/\xi=3.8(8)~{\rm GeV}$. The natural candidate for a scalar particle is the lightest glueball, $m_{0^{++}}=3.405(21)\sqrt{\sigma_0}=1.65(2)~{\rm GeV}$~\cite{Athenodorou:2020ani}, so the extracted core mass lies above the glueball mass by a factor of 1.4(2) and the scalar mass by a factor of 2.3(5). It should be noted that the identification of $\lambda$ and $\xi$ with a vector and scalar particle respectively depends on specific assumptions about the functional form of the QCD vacuum potential, and the dual superconductor picture does not rely on these specific assumptions.

We also made a scan in $\xi_v$ and $\lambda$ respectively, where we held each parameter fixed and fitted the remaining five parameters to the HERA data. We found that the incoherent cross section is very sensitive to the variation of these parameters, and at the 95\% confidence level ($\Delta\chi^2_{\rm inc.}=3.84$), $\lambda$ could only vary by 8\%, while $\xi_v$ could vary by 18\% (\ref{app:scans}).

\section{Other extracted parameters}
\label{sec:other}
The hotspot centres are distributed with $B_{c0}=3.78(15)~{\rm GeV}^{-2}$ across the best-fit range. The proton's gluon radius at $\xpom=x_0=0.0012$ then becomes $b_{\rm rms}=0.650(11)~{\rm fm}$ (\ref{app:radii}) which can be compared with the proton's transverse charge radius, $0.6261(6)$~fm, obtained via the Dirac form factor (\ref{app:radii}) from the CODATA charge radius~\cite{Mohr:2024kco}.

We may interpret the fitted saturation-scale fluctuations in terms of gluon occupancy. At $N_{hs}=8$ the fit gives $s={\rm Var}[\zeta_i]/\braket{\zeta_i}^2\approx1.4\dots1.8$ for small $|t|$, with $s$ defined in Eq.~\eqref{eq:satshare}. Across the best-fit range, $\sigma_S$ grows with $N_{hs}$ such that $s/N_{hs}\approx0.18$ stays constant. Since the hotspots fluctuate independently, $s/N_{hs}$ is the relative variance of the proton's \emph{total} small-$x$ gluon content. For a Poisson distribution the relative variance is the inverse of the mean, so the fit corresponds to $N_{hs}/s\approx5$ gluons in the proton, or $0.5\dots0.7$ per hotspot at $N_{hs}=8$. This agrees with an independent estimate from the gluon density of the dipole amplitude, $xg(x=10^{-3},\mu^2=2~{\rm GeV}^2)\approx5.0$ gluons per unit rapidity. We therefore interpret the saturation-scale fluctuations as shot noise in the number of small-$x$ gluons, which dominates the incoherent cross section at small $|t|$.

The shrinkage parameter $\alpha'_{\rm eff.}=0.060(27)_{\rm fit}(1)_{N_{hs}}~{\rm GeV}^{-2}$ is stable across all fits and uncorrelated with the geometry (absolute correlations $\le0.05$). This is smaller than the values reported from ZEUS ($\alpha'=0.115\pm 0.018^{+0.008}_{-0.015}~{\rm GeV}^{-2}$) \cite{Chekanov:2002xi} and H1 ($\alpha'=0.164 \pm 0.028 \pm 0.030~{\rm GeV}^{-2}$)~\cite{Aktas:2005xu}. 

That the data require a core with an exponential halo, rather than accommodating any two-scale profile, is shown in \ref{app:profiles} and Table~\ref{tab:EMprofiles}. Among such profiles, Clem fits at least as well as the others, but the differences are below one unit of $\chi^2$. 
Neither length is an artefact of the dipole probe: both are fixed within half their uncertainties by the data at $|t|\le8~{\rm GeV}^2$, and move by less than that for $m_c=1.318\pm0.1$~GeV (\ref{app:tcutmc}).

\section{Conclusions and outlook}
\label{sec:conclusions}
We have shown that the transverse profile of the proton's small-$x$ gluons, measured in exclusive $J/\psi$ photoproduction, is well described by the Clem vortex profile of a dual superconductor. The fit gives a penetration depth $\lambda=0.19(2)$~fm, comparable to lattice fits, a core size $\xi_v=0.087(14)$~fm and a Ginzburg--Landau parameter $\kappa=2.6(8)$, in the type-II regime. The vector mass $1/\lambda$ agrees with the lattice gluelump mass, while $1/\xi_v$ and the scalar mass $\sqrt2/\xi$ exceed the glueball mass by factors of 1.4(2) and 2.3(5). The implied GL line tension reproduces the QCD string tension for $\alpha_S=0.37(4)$, close to $\alpha_S(1/\lambda^2)=0.36(2)$ in the LO scheme of the dipole amplitude. The profile alone cannot distinguish a vortex cross section from a projected three-dimensional screened source with a core of the same size. What is specific to the vortex picture is the relation between $\xi_v$, $\lambda$ and $\kappa$, which the string tension tests.

The geometry is stable under all the variations we tested, namely the charm mass, the largest $|t|$ included in the fit and the dipole parameter set, and shows no $\xpom$-dependence, consistent with it being set by the vacuum's response to the colour sources in the target. These tests were made possible by the analytic leading-twist cross sections of \ref{app:moments}, which also show which fluctuations dominate at each $|t|$.

We predict that these results are independent of the photon virtuality $Q^2$, which measurements at the future Electron-Ion Collider or the LHeC~\cite{Accardi:2012qut,AbdulKhalek:2021gbh,LHeC:2020van} will be able to confirm, since the hotspot geometry is a property of the target while $Q^2$ controls only the resolution of the probe. If $\lambda$ and $\xi_v$ are properties of the vacuum, the same values should describe the large-$|t|$ spectrum in photoproduction of light vector mesons measured at HERA, and incoherent $J/\psi$ production off nuclei in ultra-peripheral collisions. These tests require extending the model beyond the leading-twist approximation and are left to future work.

Another prediction is that in a type-II superconductor vortex lines repel at the scale $\lambda$. We tested this by introducing a hotspot--hotspot repulsion and fitting its strength and range (\ref{app:corr}), but the data can neither confirm nor exclude it. However, at $|t|\sim1/\lambda^2\sim1~{\rm GeV}^2$, where the repulsion is expected to show up, about two thirds of the incoherent cross section comes from saturation-scale fluctuations (see Eq.~\eqref{eq:satshare}), which are degenerate with the geometric variance there. Tagging the proton dissociation mass $M_Y$ would break this degeneracy, since high-$M_Y$ events are dominated by occupancy fluctuations while low-$M_Y$ events are dominated by the longer-wavelength geometric modes. This is within reach of the far-forward instrumentation planned at the EIC, and at the LHC where the $t$-spectrum can be measured in ultra-peripheral $p$Pb collisions. A calculation of the incoherent $\dint\sigma/\dint t\dint M_Y$ cross section is left for future work.

\section*{Declaration of competing interest}
The authors declare that they have no known competing financial interests or personal relationships that could have appeared to influence the work reported in this paper.

\section*{Acknowledgements}
We express our gratitude to Raju Venugopalan and Bj\"orn Schenke at BNL for insightful discussions. The authors acknowledge the support from the physics department of IIT Delhi. \noindent Funding: This work was supported by the Anusandhan National Research Foundation (ANRF), Department of Science and Technology, Government of India [grant number CRG/2022/002507].

\appendix

\section{Derivation of the cross sections}
\label{app:moments}

\subsection{Moments of the thickness function}
Since the IPnonsat amplitude is linear in $T_p$, the $\vec b$-integral factorises, $\mathcal A=\mathcal D(\xpom,\Delta)\,\tilde T_p(\vec\Delta)$, with $\mathcal D=\int_0^\infty\dint r\int_0^1\dint z\,D_{rz}$ and $\tilde T_p(\vec\Delta)=\int\dint^2\vec b\,e^{-i\vec\Delta\cdot\vec b}\,T_p(\vec b)$. The cross sections are therefore:
\begin{eqnarray}
\frac{\dint\sigma_{\rm coh.}}{\dint t}&=&\mathcal K\,\big|\braket{\tilde T_p}\big|^2,~~
\frac{\dint\sigma_{\rm inc.}}{\dint t}=\mathcal K\Big[\braket{|\tilde T_p|^2}
-\big|\braket{\tilde T_p}\big|^2\Big],\nonumber\\
\mathcal K&=&\frac{\mathcal N R_g^2(1+\beta^2)}{16\pi}\,|\mathcal D|^2,
\label{eq:EMxsec}
\end{eqnarray}
where $R_g$ and $\beta=\tan(\pi\lambda_g/2)$ are the skewedness and real-part corrections~\cite{Shuvaev:1999ce,Kowalski:2006hc} and $\mathcal N$ is the overall normalisation. All dependence on the proton's geometry is contained in the two moments of $\tilde T_p$. Each hotspot is weighted by its saturation-scale factor $\zeta_i$ and centred on its position relative to the configuration's CM, $\vec b_i'=\vec b_i-\sum_k\vec b_k/N_{hs}$, so that
\begin{equation}
\tilde T_p(\vec\Delta)=\frac{\hat T_{hs}(t)}{N_{hs}}\sum_{i=1}^{N_{hs}}
\zeta_i\,e^{-i\vec\Delta\cdot\vec b_i'},
\label{eq:EMTp}
\end{equation}
where $\hat T_{hs}(t)=\int\dint^2\vec b\,e^{-i\vec\Delta\cdot\vec b}\,T_{hs}(\vec b)$ is the hotspot form factor, with $\hat T_{hs}(0)=1$. The $\vec b_i$ are independent, with density $T_c(\vec b)$ and characteristic function $\phi(\Delta)=\int\dint^2\vec b\,e^{-i\vec\Delta\cdot\vec b}\,T_c(\vec b)$. The $\zeta_i$ are independent of each other and of the positions, and log-normal with $\braket{\zeta_i}=1$ and ${\rm Var}[\ln\zeta_i]=4\lambda_g^2\sigma_S^2$, so that $\braket{\zeta_i^2}=e^{4\lambda_g^2\sigma_S^2}$. Writing $\vec b_i'=\frac{N_{hs}-1}{N_{hs}}\vec b_i-\frac{1}{N_{hs}}\sum_{k\neq i}\vec b_k$, the first moment factorises into $N_{hs}$ independent averages,
\begin{equation}
\braket{\tilde T_p}=\hat T_{hs}\,\phi\Big(\tfrac{N_{hs}-1}{N_{hs}}\Delta\Big)
\Big[\phi\Big(\tfrac{\Delta}{N_{hs}}\Big)\Big]^{N_{hs}-1}\equiv\hat T_{hs}F(t),
\label{eq:EMfirst}
\end{equation}
which is Eq.~\eqref{eq:F}. In the second moment the CM shift cancels, $\vec b_i'-\vec b_j'=\vec b_i-\vec b_j$, and the $N_{hs}$ diagonal and $N_{hs}(N_{hs}-1)$ off-diagonal terms give:
\begin{eqnarray}
\braket{|\tilde T_p|^2}&=&\frac{\hat T_{hs}^2}{N_{hs}^2}\sum_{i,j}
\braket{\zeta_i\zeta_j}\braket{e^{-i\vec\Delta\cdot(\vec b_i-\vec b_j)}}\nonumber\\
&=&\frac{\hat T_{hs}^2}{N_{hs}}\Big[e^{4\lambda_g^2\sigma_S^2}+(N_{hs}-1)G(t)\Big],
\label{eq:EMsecond}
\end{eqnarray}
with $G=|\phi|^2$. Subtracting $|\braket{\tilde T_p}|^2$ gives Eqs.~\eqref{eq:var} and \eqref{eq:vartot}. For a Gaussian $T_c$ of width $B_c$, $\phi=e^{-B_c|t|/2}$, $G=e^{-B_c|t|}$ and $F=G^{(N_{hs}-1)/2N_{hs}}$. For $|t|\gg1/B_c$, $F$ and $G$ vanish and the incoherent cross section tends to $\mathcal K\,e^{4\lambda_g^2\sigma_S^2}\,\hat T_{hs}^2/N_{hs}$: the large-$|t|$ fluctuations measure the single-hotspot form factor multiplied by saturation scale fluctuations. The Clem form factor in Eq.~\eqref{eq:Clem} follows from $\int_0^\infty b\,\dint b\,J_0(\Delta b)\,K_0\big(\sqrt{b^2+\xi_v^2}/\lambda\big) =\xi_vK_1(\xi_vQ)/Q$, normalised at $t=0$. 

From Eqs.~\eqref{eq:var} and \eqref{eq:vartot}, the share of the incoherent cross section due to saturation-scale fluctuations is
\begin{eqnarray}
\frac{{\rm Var}_{\rm sat.}}{{\rm Var}_{\rm total}}
&=&\frac{s}{s+1+(N_{hs}-1)G(t)-N_{hs}F^2(t)}, \nonumber \\
s&\equiv& e^{4\lambda_g^2\sigma_S^2}-1
\label{eq:satshare}
\end{eqnarray}
which is independent of the hotspot profile, since $\hat T_{hs}$ cancels in the ratio. At $N_{hs}=8$, $B_{c0}=3.75~{\rm GeV}^{-2}$ and $|t|=1~{\rm GeV}^2$, it ranges from 0.62 to 0.68 for $s={\rm Var}[\zeta_i]/\braket{\zeta_i}^2=1.4\dots1.8$. Saturation-scale fluctuations therefore contribute about two thirds of the incoherent cross section at the momentum transfer where hotspot repulsion would appear. For large $|t|$, the ratio becomes $s/(s+1)$.

\section{Systematic tests}
\label{app:systematics}

\subsection{Alternative hotspot profiles}
\label{app:profiles}
To test whether the data require the exponential halo, rather than accommodating any two-scale profile, we replaced the exponential halo by other tails, all at $N_{hs}=8$. We use a convolution of a Gaussian core of width $B_{hs}$ with a Yukawa $K_0$ profile falling as $\exp(- b/\lambda)$ as the baseline, which enables these variations. The Gaussian$\otimes K_0$ profile is very similar to the Clem profile, with the core size $\sqrt{2B_{hs}}$. These tests are summarised in Table~\ref{tab:EMprofiles}. Within the family $\hat T_{hs}(t)=e^{-B_{hs}|t|/2}(1+\lambda^2|t|/n)^{-n}$ the fit gives $n=1.17(21)$, consistent with the $K_0$ value $n=1$, with the same fit quality as the Clem profile despite one extra parameter. 
A stretched exponential $\hat T_{hs}(t)=e^{-B_{hs}|t|/2}e^{-\left(\lambda^2|t|\right)^\beta}$ gives $\beta=0.49(5)$ and a power law gives a fall-off of power $p=1.53(18)$, mutually consistent, but both are disfavoured by $\Delta\chi^2=30$ with one extra parameter. Removing the Gaussian core while letting the tail exponent float gives $n=1.57(7)$ and $\Delta\chi^2=2.6$ at equal parameter count; a pointlike screened source ($n=1$) gives $\Delta\chi^2\approx 200$. The data therefore require the exponential tail and favour a core, and $\lambda$ stays within $0.19\dots0.21$~fm for all these profiles that describe the data.
Taking the hotspot thickness proportional to the field energy density of the vortex, $T_{hs}\propto K_0^2(\sqrt{b^2+\xi_v^2}/\lambda)$, describes the data similarly well ($\Delta\chi^2=+0.2$ relative to Clem), with a profile decay length $\lambda/2=0.24(1)$~fm comparable to the Clem value of $\lambda$. The data therefore fix the range of the profile but not which power of the field it follows. The penetration depth would then be $\lambda=0.48(2)$~fm, giving $m_V=0.41(2)$~GeV, well below the gluelump mass, and reproducing the string tension would require $\alpha_S\simeq1.7$. The GL parameter then becomes $\kappa=5.8(7)$, with $\xi=0.082~{\rm fm}$. 

\setcounter{table}{0}
\begin{table}[h]
\centering
\begin{tabular}{lcc}
\hline
Hotspot profile & shape parameter & $\Delta\chi^2$ \\
\hline
Gaussian $\otimes\, K_0$ & $\lambda=0.209(7)~{\rm fm}$ & 0 \\
&  \!\!\!\!\!\!\!\!
$\sqrt{2B_{hs}}=0.090(3)~{\rm fm}$ & \\
Clem, Eq.~\eqref{eq:Clem}, MZ params. \cite{Mantysaari:2018nng} & $\lambda=0.193(8)~{\rm fm}$ & $-0.9$ \\
& $\xi_v=0.084(7)~{\rm fm}$ & \\
Clem, Eq.~\eqref{eq:Clem}, STU params. \cite{Sambasivam:2019gdd} & $\lambda=0.192(8)~{\rm fm}$ & $+1.8$ \\
& $\xi_v=0.084(7)~{\rm fm}$ & \\
Gaussian $\otimes\,(1+\lambda^2|t|/n)^{-n}$ & $n=1.17(21)$ & $-0.9$ \\
Gaussian $\otimes$ stretched exponential & $\beta=0.49(5)$ & $+30$ \\
Gaussian $\otimes$ power law & $p=1.53(18)$ & $+30$ \\
$(1+\lambda^2|t|/n)^{-n}$, no core & $n=1.57(7)$ & $+2.6$ \\
$T_{hs}\propto K_0^2\left(\sqrt{b^2+\xi_v^2}/\lambda\right)$ & $\lambda/2=0.24(1)~{\rm fm}$  & $-0.7$ \\
                    & $\xi_v=0.107(7)~{\rm fm}$ & \\
\hline
\end{tabular}
\caption{Fit quality of alternative hotspot profiles at $N_{hs}=8$, relative to the Gaussian $\otimes K_0$ fit ($\chi^2/{\rm ndf}=71.9/98$). Models with a shape parameter have one more parameter than the reference, except for the no-core and $K_0^2$ models which have the same number.}
\label{tab:EMprofiles}
\end{table}

\subsection{Parameter scans}
\label{app:scans}
Holding $\xi_v$ or $\lambda$ fixed and refitting the remaining five parameters, the coherent data are insensitive to $\xi_v$ throughout the scanned range and to the penetration depth in the range $\lambda=0.04\dots0.26~{\rm fm}$. The incoherent data, at 95\% confidence ($\Delta\chi^2_{\rm inc.}=3.84$), give $\xi_v=0.084^{+0.016}_{-0.013}$~fm and $\lambda=0.193^{+0.015}_{-0.016}$~fm. The shape parameters are therefore fixed by the incoherent data.

\subsection{Energy dependence}
\label{app:energy}

Letting the Gaussian hotspot core width evolve as $B_{hs}=B_{hs0}+2\beta'\log(x_0/\xpom)$ gives $\beta'=0.016(16)~{\rm GeV}^{-2}$, consistent with zero, and improves the fit by $\Delta\chi^2=-1.25$ for one extra parameter. Letting the hotspot number grow as $\braket{N_{hs}}=N_{hs0}(x_0/\xpom)^{\delta_N}$ (similar to \cite{Cepila:2016uku}) gives $\delta_N=0.04(11)$, i.e.\ $\braket{N_{hs}}=7.3\dots8.8$ over $\xpom=10^{-2}\dots10^{-4}$, and $\Delta\chi^2=-0.06$, with one extra parameter. In both variants $\alpha'_{\rm eff.}=0.060(27)~{\rm GeV}^{-2}$ is unchanged. 

We can also add an energy dependence directly to the Clem parameters, either $\xi_v=\xi_{v0}+2\beta'_\xi\log(x_0/\xpom)$ or $\lambda=\lambda_0[1+\delta_\lambda\log(x_0/x)]$, where $x=M_{J/\psi}^2/W^2$ is $\xpom$ at $t=0$. We find $\beta'_\xi=0.007(31)~{\rm GeV}^{-1}$ and $\delta_\lambda=-0.02(4)$, both consistent with zero. A hotspot whose size scales with the inverse saturation scale, $Q_s^2\propto\xpom^{-\lambda_g}$, would instead give $\delta_\lambda\approx-\lambda_g/2\approx-0.09$, which is disfavoured at about $2\sigma$. 

\subsection{The $t$-cut and the charm mass}
\label{app:tcutmc}
To further test that the Clem parameters are not a remnant of the probe, we scan them as a function of the maximum $|t|$ included in the fit. The large-$|t|$ part of the spectrum could be affected by the finite size of the dipole probe. Restricting the fit to $|t|\le6~(4)~{\rm GeV}^2$ gives $\lambda=0.200(13)~(0.206(19))$~fm and $\xi_v=0.072(18)~(0.058(33))$~fm, consistent with the full fit. For any $t_{\max}\ge8~{\rm GeV}^2$ all parameters agree with the full fit to within half their uncertainties. The hotspot geometry is therefore determined by the range $4\leq|t|\le 8~{\rm GeV}^2$; the fit still describes the large-$|t|$ spectrum up to $|t|=30~{\rm GeV}^2$. 

We also vary the charm mass to check that the core size, which is of similar magnitude, is not a remnant of this scale. Varying the charm-quark mass in the dipole amplitude from 0.75 to 1.5 times its nominal value $m_c=1.318$~GeV changes $\lambda$ by less than 1\%, while $\xi_v$ moves from 0.075 to 0.093~fm and $\kappa$ from 3.1 to 2.4. The core therefore does not track the size of the probe, which would make it shrink as $m_c$ grows, but it partly compensates the probe's $t$-dependence. For $m_c$ within $\pm0.1$~GeV of its nominal value the shifts are below half the fit uncertainties.

\subsection{Hotspot correlations}
\label{app:corr}
To test for hotspot repulsion, we weighted the joint distribution of the hotspot centres by a factor $1-c\,h(\vec b_i-\vec b_j)$ for every pair, with a Gaussian hole $h(\vec s)=e^{-s^2/2r_c^2}$ of range $r_c$, where $c>0$ gives repulsion and $c<0$ attraction, and derived $F(t)$ and $G(t)$ consistently from the same reweighted distribution. With the Clem profile at $N_{hs}=8$, freeing $c$ and $r_c$ changes $\chi^2$ by less than 0.11 for two additional parameters, and for any fixed $c$ between $-0.6$ and 3 the fit gives the same $\chi^2$ by shrinking $r_c$ until the hole no longer matters. At fixed $r_c$ the refitted strength stays below $|c|=0.22$ for $0.1\le r_c\le2$~fm, with $c=0.10$ at $r_c=0.2~{\rm fm}\simeq\lambda$, where a repulsion of $c=0.35$ worsens the fit by $\Delta\chi^2=1$, mainly through the coherent data. This limit assumes Gaussian-distributed centres and may weaken for a flatter distribution. Neither repulsion nor attraction is preferred, and up to this limit $\lambda$ and $\xi_v$ shift by less than 0.6 of their uncertainties. HERA data cannot resolve the correlation because at $|t|\sim1/\lambda^2$ about two thirds of the incoherent cross section comes from saturation-scale fluctuations (Eq.~\eqref{eq:satshare}), which do not depend on where the hotspots are.
\section{The string tension}
\label{app:string}
Clem's trial vortex~\cite{Clem:1975ohd} has the order-parameter modulus $f=\rho/\sqrt{\rho^2+\xi_v^2}$ and the field profile in Eq.~\eqref{eq:Clem}. Its Ginzburg--Landau energy per unit length, Eqs.~(10)--(15) in Ref.~\cite{Clem:1975ohd}, is, for a vortex with flux $\Phi$ in natural units:
\begin{eqnarray}
\varepsilon&=&\frac{\Phi^2}{4\pi\lambda^2}\,L \label{eq:fluxE}\\
L&=&2\kappa H_{c1}'=\frac{\kappa^2\alpha^2}{4}+\frac14
+\frac{K_0(\alpha)}{\alpha K_1(\alpha)},
\label{eq:EMeps}
\end{eqnarray}
where $H_{c1}'$ is Clem's reduced lower critical field. Minimising in $\alpha$ gives Clem's Eq.~(16), our Eq.~\eqref{eq:GL}, $\kappa^2\alpha^2=2[1-K_0^2(\alpha)/K_1^2(\alpha)]$, and eliminating $\kappa$,
\begin{equation}
L(\alpha)=\frac34-\frac{K_0^2(\alpha)}{2K_1^2(\alpha)}+\frac{K_0(\alpha)}{\alpha K_1(\alpha)}.
\label{eq:EML}
\end{equation}
At $\kappa=1/\sqrt2$ this gives $\xi_v/\xi=0.935$ and $H_{c1}'=0.732$, as quoted by Clem, which gives $L=1.035$ compared to the exact GL value $L=1$.

In the dual GL theory of Ref.~\cite{Koma:2002cv}, with $eg=4\pi$, the
$N_c(N_c-1)/2$ dual gauge fields enter with magnetic coupling $g_m=\sqrt{N_c/2}\,g$, so a unit winding gives $\Phi=2\pi/g_m$.
A quark of colour $k$ couples to $N_c-1$ condensates, and its flux tube consists of one elementary vortex in each and their energies add. Plugging into Eq.~\eqref{eq:fluxE} gives
\begin{equation}
\sigma=(N_c-1)\frac{(2\pi/g_m)^2}{4\pi\lambda^2}L
=\frac{N_c-1}{2N_c}\,\frac{e^2}{4\pi}\,\frac{L}{\lambda^2}=C\alpha_S\frac{L}{\lambda^2} \nonumber
\end{equation}
with $C=1/3$ for $N_c=3$. With $\lambda=1/m_B=1/(\sqrt{N_c}\,gv)$ this is
$\sigma=2\pi(N_c-1)v^2L$, which for $L=1$ is the Bogomol'nyi value,
Eq.~(4.6) of Ref.~\cite{Koma:2002cv}. The fitted $\alpha$ gives $L=1.79(22)$.

\section{Gluon and charge radii}
\label{app:radii}
We define the gluon radius from the first moment of the proton thickness,
$b_{\rm rms}^2=-4\frac{\dint\ln\braket{\tilde T_p}}{\dint|t|}\big|_{t=0}$. Eqs.~\eqref{eq:EMfirst} and \eqref{eq:Clem} give:
\begin{eqnarray}
b_{\rm rms}^2=2B_c\frac{N_{hs}-1}{N_{hs}}+4\lambda^2+2\lambda\xi_v\frac{K_0(\alpha)}{K_1(\alpha)},
\label{eq:EMbrms}
\end{eqnarray}
For $\xpom=x_0$ we get $b_{\rm rms}=0.65(1)$~fm across the best-fit range of $N_{hs}$. The electromagnetic counterpart is the transverse charge density, the two-dimensional Fourier transform of the Dirac form factor $F_1$, measured from the transverse centre of momentum~\cite{Burkardt:2002hr}. Expanding $F_1(Q^2)=\int\dint^2\vec b\,\rho(b)\,e^{i\vec q\cdot\vec b}=1-Q^2\braket{b^2}_{\rm ch}/4+\dots$ gives $\braket{b^2}_{\rm ch}=-4\frac{\dint F_1}{\dint Q^2}\bigg|_{Q^2=0}$. Following \cite{Gramolin:2021gln}, the Sachs form factor is $G_E=F_1-(Q^2/4m_p^2)F_2$, $F_2(0)=\mu_p-1$ where the proton's magnetic moment in nuclear magnetons $\mu_p\approx 2.79$, and $r_E^2=-6\frac{\dint G_E}{\dint Q^2}\big|_{Q^2=0}$, we get:
\begin{equation}
\braket{b^2}_{\rm ch}=\tfrac23\,r_E^2-\frac{\mu_p-1}{m_p^2},
\label{eq:EMbch}
\end{equation}
The CODATA value $r_E=0.84075(64)$~fm~\cite{Mohr:2024kco} gives $\sqrt{\braket{b^2}_{\rm ch}}=0.6261(6)$~fm.

\section*{Data availability}
The HERA data used in this work are publicly available in Refs.~\cite{Aktas:2005xu,Alexa:2013xxa,Aktas:2003zi, Chekanov:2002rm,Chekanov:2009ab}. The fitting code is available at \url{https://github.com/tobiastoll/sartre/tree/ClemIPnonsatFits}

\section*{Declaration of generative AI and AI-assisted technologies in the manuscript preparation process}
During the preparation of this work the authors used Claude (Anthropic) to assist with writing and debugging the fitting code, and to cross-check derivations and numerical results. After using this tool, the authors reviewed and edited the content as needed and take full responsibility for the content of the published article.
\bibliographystyle{elsarticle-num}
\bibliography{Saturationandfluctuations}

\end{document}